\documentclass[pra,twocolumn,tightenlines,showpacs,nofootinbib,groupedaddress]{revtex4}
\usepackage{bm,dcolumn,graphicx}

\begin{document}
\title{High-accuracy relativistic many-body calculations of van der Waals  coefficients,
$C_6$, for alkaline-earth atoms}
\author{Sergey G. Porsev}
\altaffiliation{Permanent Address:
Petersburg Nuclear Physics Institute, Gatchina,
Leningrad district, 188300, Russia.}
\affiliation{Physics Department, University of Nevada, Reno,
Nevada 89557-0058.}
\author{Andrei Derevianko}
\affiliation{Physics Department, University of Nevada, Reno,
Nevada 89557-0058.}

\date{\today}

\begin{abstract}
Relativistic many-body calculations of van der Waals  coefficients
$C_6$ for dimers correlating to two ground state alkaline-earth atoms
at large internuclear separations are reported.
The following values and uncertainties were determined :
$C_6 = 214(3)$ for Be, 627(12) for Mg, 2221(15) for Ca, 3170(196) for Sr, and
5160(74) for Ba in atomic units.
\end{abstract}


\maketitle
\paragraph{Introduction.}
The realization of Bose-Einstein
con\-densation (BEC) in dilute ultracold samples of hydrogen and alkali-metal atoms
Li, Na, and Rb~\cite{IngStrWie99} has prompted a search for other atomic
and molecular species where BEC can  possibly be attained.
Of non-alkali atoms, so far the condensation was  successful only with
metastable helium~\cite{SanLeoWan01,RobSirBro01}. Cooling and trapping
experiments with alkaline-earth atoms (in particular Mg, Ca, and Sr) were recently
reported (see, e.g., Ref.~\cite{ZinBinRie00,KatIdoIso99,DinVogHal99}) and
prospects of achieving the condensation with alkaline-earth atoms were also
discussed~\cite{MacJulSuo99,ZinBinRie00}. Ultracold alkaline--earth atoms
possess several advantages over alkali-metal atoms.
For example, utilization of the narrow
spin-forbidden transition $\,^1\!S_0 \rightarrow \,^3\!P_1^o$ permits to optically
cool atoms down to  nano--Kelvin regime \cite{KatIdoIso99,WalErt89}.
There is also a number of isotopes available with zero
nuclear spin, so that the resulting molecular potentials
are not complicated by the ``spaghetti'' of hyperfine-structure
states; this simplifies studies of trap losses and ultracold collisions~\cite{Jul00}.

Here we apply relativistic many-body methods
to the determination of dispersion (van der Waals) coefficients $C_6$
for the interaction of two identical alkaline-earth atoms in their respective ground
states. The leading interaction of two ground-state atoms at large internuclear
separations $R$ is parameterized as $-C_6/R^6$. Knowledge
of the dispersion coefficients is required, for example, in determination
of scattering lengths governing
properties of Bose-Einstein condensates
of dilute samples~\cite{WeiBagZil99}.

We employ several atomic relativistic many-body methods of varying accuracy.
The dominant contribution to $C_6$ was evaluated with the configuration
interaction (CI) method coupled with many-body perturbation
theory (MBPT)~\cite{DzuFlaKoz96P,DzuKozPor98}; smaller
terms were computed using the less accurate relativistic random-phase
approximation (RRPA)  and Dirac-Hartree-Fock (DHF) methods.
The values were further adjusted with accurate theoretical
and experimental data for electric-dipole matrix elements
and energies of principal transitions.
We tabulate
the values of $C_6$ for Be, Mg, Ca, Sr, and Ba. We also estimate
uncertainties to be on the order of 1-2\% for
all alkaline-earth atoms, except for a 5\% accuracy for Sr.

\paragraph{Method of calculations.}
The dispersion coefficient $C_6$ describes a second-order
correction to energies of molecular terms related to the
interaction of atomic dipoles induced by molecular fields.
Due to large internuclear separations involved,
the two-center molecular-structure problem
is reduced to the determination of {\em atomic} properties.
The van der Waals coefficient
may be expressed as~\cite{DalDav66}
\begin{equation}
 C_6  = 6 \sum_{ij}
 \frac{ |\langle \Psi_g |D_z | \Psi_i \rangle|^2 |\langle \Psi_g |D_z | \Psi_j \rangle|^2  }
 {(E_i -E_g) +(E_j -E_g) } \, , \label{Eqn_C6_direct}
\end{equation}
where $\Psi_g$ and $E_g$ are the wavefunction and energy of the atomic ground state,
$D_z$ is an electric-dipole operator, and the summation
is over intermediate atomic states $\Psi_i$ and $\Psi_j$ with respective
energies $E_i$ and $E_j$.
Atomic units $\hbar=|e|=m_e=1$ are used throughout.
The above relation can be recast into the  Casimir-Polder form
\begin{equation}
 C_6 = \frac{3}{\pi} \int_0^\infty [\alpha(i\omega)]^2 \,
d \omega \, , \label{Eqn_C6_polariz}
\end{equation}
where $\alpha(i \omega )$ is the dynamic polarizability of
imaginary argument defined as
\begin{equation}
\alpha ( i \omega ) = 2 \mathrm{Re} \sum_i
\frac{ \langle \Psi_g |  D_z | \Psi_i \rangle \langle
\Psi_i | D_z | \Psi_g \rangle}
{ \left( E_i-E_g \right) + i \omega } \, . \label{Eqn_alpha}
\end{equation}

The intermediate states in the sum, Eq.~(\ref{Eqn_alpha}), can be separated into
valence and core-excited states. We write
\begin{equation}
\alpha ( i \omega ) = \alpha_v ( i \omega ) + \alpha_c ( i \omega ) +
\alpha_{cv} ( i \omega ) \, .
\label{Eqn_alphaBreak}
\end{equation}
To determine the valence contribution $\alpha_v$
we employ combined relativistic configuration interaction method and many-body perturbation theory
(CI+MBPT).
Smaller contributions of core-excited states $\alpha_c$ are estimated with
the relativistic random-phase approximation for the atomic core. In this method excitations
of core electrons are allowed into the occupied valence shell and we introduce
the correction $\alpha_{cv}$ to account for
a subsequent violation of the Pauli exclusion principle; this small correction
is evaluated using the Dirac-Hartree-Fock method.

Similar relativistic many-body techniques were involved in our previous high-precision
determination of van der Waals coefficients for atoms with one valence
electron outside a closed core~\cite{DerJohSaf99,DerBabDal01}. Divalent atoms, considered
here, present an additional challenge due to a strong Coulomb
repulsion of the valence electrons. This strong interaction
is treated here with the configuration interaction method and smaller
residual corrections (like core polarization) are treated with the many-body
perturbation theory. The method, designated as CI+MBPT, was
developed in
Ref.~\cite{DzuFlaKoz96P,DzuFlaKoz96Z,DzuKozPor98,PorRakKoz99J,PorRakKoz99P,KozPor99E}.

Here we briefly recap the main features of the CI+MBPT method.
The complete functional space for electronic wavefunctions is partitioned in
two parts: the model space spanning all possible excitations
of the two valence electrons  and an orthogonal
space which adds various excitations of core electrons.
The valence CI basis set is saturated; e.g., the Ba ground state
wavefunction is
represented as a combination of 1450 relativistic configurations in our calculations.
Application of perturbation theory
leads to effective operators encapsulating many-body effects and acting
in the model space.
For example, the CI valence wavefunctions are determined from the
Schr\"odinger equation
\begin{equation}
\label{H}
H_{\rm eff}(E_n) \, | \Psi_n \rangle = E_n \, |\Psi_n \rangle \, ,  
\end{equation}
with the effective Hamiltonian defined as
\begin{equation}
\label{Heff}                                                         
  H_{\rm eff}(E) = H_{0} + C + \Sigma(E).
\end{equation}
Here $H_{0}$ is the lowest-order Dirac-Fock Hamiltonian,
$C$ is the residual Coulomb interaction between valence electrons,
and $\Sigma$ is the energy-dependent
self-energy operator corresponding to core-polarization
effects in model-potential approaches. The operator $\Sigma$
completely accounts for the second order of perturbation theory.
By the same virtue, one introduces
an effective (dressed) electric-dipole operator $D_{\rm eff}$
acting in the model space.
We determine this effective operator using
the random-phase approximation (RPA)~\cite{PorKozRak00Z,PorKozRak01}.
Qualitatively, the RPA
describes a shielding of the externally applied field by the core
electrons.

The dynamic {\em valence} polarizability $\alpha_v(i \omega)$ was computed with
the Sternheimer~\cite{Ste50} or Dalgarno-Lewis \cite{DalLew55} method
implemented in  the  CI+MBPT+RPA framework.
At the heart of the method is a solution of an
inhomogeneous Schr\"odinger equation
for a ``perturbed'' state $| \delta \Psi_\omega \rangle$
\begin{equation}
(H_{\rm eff} - E_g + i \omega)
 | \delta \Psi_\omega \rangle = (D_z)_{\rm eff} |\Psi_g \rangle \, ,
\label{Eqn_inhom}
\end{equation}
so that
\begin{equation}
\alpha_v (i \omega) = 2 \mathrm{Re} \langle \Psi_g | (D_z)_{\rm eff}
| \delta \Psi_\omega \rangle \, .
\end{equation}
In these expressions the electric-dipole operator $D_{\rm eff}$ is calculated
at the CI+MBPT+RPA level of approximation.
Present approach is a frequency-dependent generalization of calculations
of static dipole polarizabilities reported in~\cite{KozPor99E,JohChe96};
technical details can be found in these works.

The overwhelming contribution (on the order of 90\%) to the value of the
van der Waals coefficient, Eq.~(\ref{Eqn_C6_direct}), comes from the
lowest-energy excited $nsnp\,^1\!P^o_1$ state. Therefore the
calculated $C_6$ are mostly sensitive to accuracies of
dipole matrix elements and energy separations of the principal
$nsnp\,^1\!P^o_1 - ns^2\,^1\!S_0$
transitions. We explicitly calculated these quantities using
the same level of CI+MBPT+RPA approximation as employed in the
solution of the inhomogeneous equation~(\ref{Eqn_inhom}); these values
are marked as CI+MBPT+RPA and CI+MBPT in Table~\ref{Tab_princComp}.
We find a good agreement
with more sophisticated {\em ab initio}~\cite{PorKozRak00Z,PorKozRak01}
and experimental
values~\cite{ZinBinRie00,BizHub90,Moo58,NIST_ASD} (see Table~\ref{Tab_princComp}.)
For Be we also computed additional many-body corrections;
they can be neglected at the level of the quoted significant figures in Table~\ref{Tab_princComp}.
We conservatively
estimated an uncertainty in the matrix element for Be as a half of the
difference between valence CI and correlated value.

\begin{table}
\caption{Reduced matrix elements $D$
and energy separations $\Delta E_p$  for transitions from the
lowest-energy $nsnp\,^1\!P^o_1$ to the ground $ns^2 \,^1\!S_0$ state.
}
\label{Tab_princComp}
\begin{ruledtabular}
\begin{tabular}{ldlll}
& \multicolumn{2}{c}{$D$ }&
       \multicolumn{2}{c}{ $\Delta E_p $ }   \\
        \cline{2-3}  \cline{4-5}
      & \multicolumn{1}{c}{ CI+MBPT+RPA }    &  \multicolumn{1}{c}{accurate} &
        \multicolumn{1}{c}{ CI+MBPT}    &  \multicolumn{1}{c}{Expt.\footnotemark[1]} \\
\hline
Be & 3.26  & 3.26(1)\footnotemark[2] &0.194291 &0.193942 \\
Mg & 4.03 & 4.03(2)\footnotemark[3]  &0.159173 &0.159705\\
Ca & 4.93 & 4.967(9)\footnotemark[4] &0.107776 &0.107768\\
Sr & 5.31 & 5.28(9)\footnotemark[3]  &0.098508 &0.098866\\
Ba & 5.52  & 5.466(23)\footnotemark[5]&0.082891&0.082289 \\
\end{tabular}
\footnotemark[1]{Ref.~\cite{Moo58,NIST_ASD},}
\footnotemark[2]{This work,}
\footnotemark[3]{Ref.~\cite{PorKozRak00Z,PorKozRak01},}
\footnotemark[4]{Ref.~\cite{ZinBinRie00}.}
\footnotemark[5]{Ref.~\cite{BizHub90}.}
\end{ruledtabular}
\end{table}

Due to the enhanced sensitivity of $C_6$ to uncertainties in
the dipole matrix element and the energy separation $\Delta E_p$
of principal transitions,
we further correct the calculated dynamic polarizability by
subtracting the {\em ab initio} CI+MBPT+RPA contribution of the principal transition
\begin{equation}
\alpha_p(i \omega) = \frac{2}{3}
 \frac{\Delta E_p }
 { (\Delta E_p)^2 + \omega^2}  \, |\langle ns^2 \,^1\!S_0 ||D|| nsnp \, ^1\!P^o_1 \rangle|^2
\label{Eqn_alpPrinc}
\end{equation}
from $\alpha(i \omega)$ and adding it back with experimental energies and high-accuracy matrix elements
compiled in Table~~\ref{Tab_princComp}.

The ``perturbed'' state $| \delta \Psi_\omega \rangle$ in Eq.~(\ref{Eqn_inhom})
is defined in the model space of the valence electrons, i.e., it is
comprised from all possible valence excitations  from the
ground state $|\Psi_g \rangle $.  Since the core-excited states do not enter
the model space, their contribution to the polarizability has to be added
separately. Here we follow our work~\cite{DerJohSaf99} and use
the relativistic random-phase
approximation~\cite{Joh88} to determine the dynamic core polarizability as
\begin{equation}
 \alpha_c( i \omega) =
 \sum_{\omega_\mu > 0}
\frac{ f_\mu }
{ \left( \omega_\mu \right)^2 + \omega^2 } \, . \label{Eqn_acRRPA}
\end{equation}
Here the summation is over particle-hole excitations
from the ground state of the atomic core; $\omega_\mu$ are excitation energies
and $f_\mu$ are the corresponding electric-dipole oscillator strengths.
Accounting for core excitations is
essential in our accurate calculations, especially for heavier
atoms. For example, for Ba they contribute as much as 15\%
to the total value of $C_6$.

The particle-hole excitations summed over in Eq.~(\ref{Eqn_acRRPA})
include Pauli-principle violating excitations into the occupied valence
shell. We explicitly subtract their contribution; this small correction
$\alpha_{cv} (i \omega)$ is computed with the Dirac-Hartree-Fock method.

Our calculated dynamic polarizabilities
satisfy two important relations: (i) $\alpha(\omega=0)$ is the  ground-state
static dipole polarizability  and (ii) as a consequence of the
nonrelativistic Thomas-Reiche-Kuhn  sum rule, at
large frequencies $\omega^2 \, \alpha (i\omega ) \rightarrow N $,
where $N$ is the total number of atomic electrons.
Indeed, for Ca we obtain $\alpha(0)=160$ a.u.,
while the experimental value~\cite{MilBed76} is 169(17) a.u.
For Sr we obtain 199 a.u. which is in agreement with the measured
value~\cite{SchMilBed74} of 186(15) a.u.
And, finally, for Ba the computed
static polarizability of  273 a.u. also compares well with experimental
value~\cite{SchMilBed74} of 268(22) a.u.
Similarly, at large $\omega$, in our  calculations
the product $\omega^2 \, \alpha (i\omega )$ approaches 3.99 for Be, 11.9 for
Mg, 19.71 for Ca, 37.1 for Sr, and 54.01 for Ba;
these asymptotic {\em relativistic} values  are slightly
smaller than the exact {\em nonrelativistic} limits.
\paragraph{Results and theoretical uncertainties.}
We combine  various parts of the dynamic polarizability,
Eq.~(\ref{Eqn_alphaBreak}),
and then obtain dispersion coefficients $C_6$
with a quadrature, Eq.~(\ref{Eqn_C6_polariz}).
The resulting values of van der Waals coefficients
are presented in Table~\ref{Tab_C6fin_comp}. In this
Table, values marked
{\em ab initio} were determined in the relativistic CI+MBPT+RPA  framework.
The values marked final are {\em ab initio} values adjusted for accurate dipole matrix elements
and energies of principal transitions, compiled in Table~\protect\ref{Tab_princComp}.

\begin{table}
\caption{van der Waals coefficients $C_6$  for dimers correlating
to ground states of alkaline-earth atoms in a.u. Values marked
{\em ab initio} were determined in the relativistic CI+MBPT+RPA  framework.
The values marked final are {\em ab initio} values adjusted for accurate dipole matrix elements
and energies of principal transitions, compiled in Table~\protect\ref{Tab_princComp}.
\label{Tab_C6fin_comp}}
\begin{ruledtabular}
\begin{tabular}{llllll}
                         &
\multicolumn{1}{c}{ Be}  &
\multicolumn{1}{c}{ Mg}  &
\multicolumn{1}{c}{ Ca}  &
\multicolumn{1}{c}{ Sr}  &
\multicolumn{1}{c}{ Ba}  \\
\hline
{\em Ab initio}             & 213     & 631     & 2168     & 3240      & 5303\\
Final       & 214(3) & 627(12)  & 2221(15) & 3170(196) & 5160(74)\\[3pt]
\multicolumn{6}{c}{Other works} \\
\protect\citet{Sta94}       & 216 & 648   & 2042 & 3212 & \\
S\&C~\protect\cite{StaCer85}& 220 & 634   & 2785 &  \\
M\&K~\protect\cite{MaeKut79}& 208 & 618   & 2005 \\
A\&Ch~\protect\cite{AmuChe75}&254 &       & 2370 \\
\protect\citet{Stw71}       &    & 683(35)& \\
\end{tabular}
\end{ruledtabular}
\end{table}

Different classes of intermediate states in Eq.~(\ref{Eqn_alpha})
contribute at drastically different levels to the total values
of dispersion coefficients. For example, for Ca, the principal
$4s4p\,^1\!P^o_1 - 4s^2\,^1\!S_0$ transition contributes 85\% to the
values of $C_6$, remaining valence-valence excitations contribute 8\%,
core-excited states contribute 8\% and the counter term $\alpha_{cv}$  modifies
the final result only by -0.4\%. To estimate
dominant theoretical uncertainties we  approximate $C_6$ as
\begin{eqnarray}
 C_6 &\approx& \frac{3}{\pi} \int_0^\infty [\alpha_p (i \omega)]^2 d\omega +
 \frac{6}{\pi} \int_0^\infty \alpha_p (i \omega) \alpha_r (i \omega) d\omega
\nonumber \\
&=& C_6^{pp} + C_6^{pr} \, . \label{Eqn_C6errest}
\end{eqnarray}
Here $\alpha_p$ is a contribution of the principal transition Eq.~(\ref{Eqn_alpPrinc}),
and $\alpha_r = \alpha_v' + \alpha_c$ is a contribution of the remaining valence
states ($\alpha_v'=\alpha_v - \alpha_p$) and core-excited states. From a direct calculation for Ca
we find that this approximation recovers 99.3\% of the $C_6$ obtained from the full
expression~(\ref{Eqn_C6_polariz}).
Based on Eq.~(\ref{Eqn_C6errest}) the sensitivity of $C_6$ to
uncertainties $\delta D$
in the matrix element $D$ of the principal transition is
\begin{equation}
\delta_D C_6 \approx \left( 4 C_6^{pp} + 2 C_6^{pr} \right) \frac{ \delta D} {D} \, .
\end{equation}
To evaluate the sensitivity of $C_6$ to uncertainties in the residual polarizability
we follow Ref.~\cite{Der01a}.
In the second term of Eq.~(\ref{Eqn_C6errest}) a narrow function
$\alpha_p(i \omega)$ is integrated with a relatively
broad distribution $\alpha_r (i \omega)$. Therefore we can approximate that
\begin{equation}
\int_0^\infty \alpha_p (i \omega) \alpha_r (i \omega) d\omega \approx
\alpha_r (0) \int_0^\infty \alpha_p (i \omega)  d\omega
\end{equation}
and the sensitivity of $C_6$ is
\begin{equation}
\delta_{\alpha_r} C_6 \approx C_6^{pr} \frac{ \delta \alpha_r(0)}{\alpha_r(0)} \, .
\end{equation}
The uncertainty in the  residual  static
polarizability $\delta \alpha_r(0)$ is
a sum of uncertainties   in the contributions
of valence states beyond principal transition $\delta \alpha_v'(0)$ and
core-excited states $\delta \alpha_c(0)$.
The RRPA static dipole core polarizabilities for alkali-metal atoms
are known~\cite{DerJohSaf99} to be in a 1\% agreement with those
deduced from semiempirical analysis of Rydberg spectra; we approximate that
$\delta \alpha_c(0) \approx 0.01 \alpha_c(0)$. Further we estimate that
$\delta \alpha_v'(0) \approx \delta \alpha_p(0)$, i.e. the difference
of the contributions of the principal transition to static polarizability calculated with
CI+MBPT+RPA and accurate values compiled in Table~\ref{Tab_princComp}.

The error bars of the final values of dispersion coefficients
in Table~\ref{Tab_C6fin_comp} were calculated by adding the uncertainties
$\delta_D C_6$ and $\delta_{\alpha_r} C_6$ in quadrature.
For all considered alkaline-earth atoms the uncertainty in $C_6$ induced
by errors in matrix elements of principal transition, $\delta_D C_6$,
dominates over $\delta_{\alpha_r} C_6$.
The estimated total uncertainties are in the order of 1-2\% for
all alkaline-earth atoms, except for Sr where the accuracy is 5\%.
Similar error analysis for alkali-metal atoms~\cite{DerJohSaf99}
has proven to be  reliable;
for example, for Cs  the predicted $C_6 =6851(74)$ a.u.
was found to be in agreement with a value~\cite{LeoWilJul00} of 6890(35) a.u.
deduced from an analysis of
magnetic-field induced Feshbach resonances and photoassociation data.
However, we emphasize
that in the case of alkali-metals a number of independent high-accuracy data
was available for the dominant principal transitions ensuring
reliability of derived dispersion coefficients. This is not the
case for alkaline-earth atoms.
In our present calculation
we rely on the quoted uncertainties of accurate dipole matrix elements listed in
Table~\ref{Tab_princComp}.

A comparison with other theoretical and semiempirical determinations is presented
in Table~\ref{Tab_C6fin_comp}. There is a reasonable agreement
among different approaches for Be and Mg; results
for Ca are less consistent due to a more significant role of correlations and
core-excited states. Coupled-cluster calculations
by \citet{Sta94} were most elaborate among theoretical
treatments. We find a good agreement with
his predictions. Unfortunately, most of the authors
do not estimate  uncertainties of their methods. One of the exceptions
is Ref.~\cite{StaCer85} where sum rules and Pade-approximants
were used to establish bounds on $C_6$. For Ca, they found $2740 \leq C_6 \leq 2830$ a.u.
However,
large uncertainties of underlying experimental
data were not included in these bounds (see also Ref.~\cite{Sta94}); this explains
a significant deviation of our prediction for Ca, $C_6=2221(15)$ a.u.,
from constraints of Ref.~\cite{StaCer85}.
\paragraph{Conclusion.}
We carried out relativistic many-body calculations of van der Waals  coefficients
$C_6$ for dimers correlating to two ground state alkaline-earth atoms
at large internuclear separations. The values were adjusted with
accurate theoretical and experimental data for the electric-dipole matrix elements
and energies of the principal transitions. It is  worth emphasizing that the
dispersion coefficients depend sensitively on electric-dipole matrix
elements of principal transitions. As more accurate data for the
matrix elements become
available, for example from photoassociation experiments with
ultracold samples, the van der Waals coefficients can be constrained
further within our many-body approach.

We would like to thank W. R. Johnson, A. Dalgarno, E. Emmons, and H. R. Sadeghpour for
helpful discussions.
The work of A.D. was partially supported by the NSF
and by the Chemical Sciences, Geosciences and Biosciences Division of the Office of Basic Energy
Sciences, Office of Science, U.S. Department of Energy.

%


\begin{thebibliography}{37}
\expandafter\ifx\csname natexlab\endcsname\relax\def\natexlab#1{#1}\fi
\expandafter\ifx\csname bibnamefont\endcsname\relax
  \def\bibnamefont#1{#1}\fi
\expandafter\ifx\csname bibfnamefont\endcsname\relax
  \def\bibfnamefont#1{#1}\fi
\expandafter\ifx\csname citenamefont\endcsname\relax
  \def\citenamefont#1{#1}\fi
\expandafter\ifx\csname url\endcsname\relax
  \def\url#1{\texttt{#1}}\fi
\expandafter\ifx\csname urlprefix\endcsname\relax\def\urlprefix{URL }\fi
\providecommand{\bibinfo}[2]{#2}
\providecommand{\eprint}[2][]{\url{#2}}

\bibitem[{\citenamefont{Inguscio et~al.}(1999)\citenamefont{Inguscio,
  Stringari, and Wieman}}]{IngStrWie99}
\bibinfo{editor}{\bibfnamefont{M.}~\bibnamefont{Inguscio}},
  \bibinfo{editor}{\bibfnamefont{S.}~\bibnamefont{Stringari}},
  \bibnamefont{and} \bibinfo{editor}{\bibfnamefont{C.}~\bibnamefont{Wieman}},
  eds., \emph{\bibinfo{title}{Bose-Einstein Condensation in Atomic Gases}},
  Proc. of the International School of Physics {``Enrico Fermi,'' Course CXL}
  (\bibinfo{publisher}{IOS Press}, \bibinfo{address}{Amsterdam},
  \bibinfo{year}{1999}).

\bibitem[{\citenamefont{Dos~Santos et~al.}(2001)\citenamefont{Dos~Santos,
  Leonard, Wang, Barrelet, Perales, Rasel, Unnikrishnan, Leduc, and
  Cohen-Tannoudji}}]{SanLeoWan01}
\bibinfo{author}{\bibfnamefont{F.~P.} \bibnamefont{Dos~Santos}},
  \bibinfo{author}{\bibfnamefont{J.}~\bibnamefont{Leonard}},
  \bibinfo{author}{\bibfnamefont{J.}~\bibnamefont{Wang}},
  \bibinfo{author}{\bibfnamefont{C.~J.} \bibnamefont{Barrelet}},
  \bibinfo{author}{\bibfnamefont{F.}~\bibnamefont{Perales}},
  \bibinfo{author}{\bibfnamefont{E.}~\bibnamefont{Rasel}},
  \bibinfo{author}{\bibfnamefont{C.~S.} \bibnamefont{Unnikrishnan}},
  \bibinfo{author}{\bibfnamefont{M.}~\bibnamefont{Leduc}}, \bibnamefont{and}
  \bibinfo{author}{\bibfnamefont{C.}~\bibnamefont{Cohen-Tannoudji}},
  \bibinfo{journal}{Phys. Rev. Lett.} \textbf{\bibinfo{volume}{86}},
  \bibinfo{pages}{3459} (\bibinfo{year}{2001}).

\bibitem[{\citenamefont{Robert et~al.}(2001)\citenamefont{Robert, Sirjean,
  Browaeys, Poupard, Nowak, Boiron, Westbrook, and Aspect}}]{RobSirBro01}
\bibinfo{author}{\bibfnamefont{A.}~\bibnamefont{Robert}},
  \bibinfo{author}{\bibfnamefont{O.}~\bibnamefont{Sirjean}},
  \bibinfo{author}{\bibfnamefont{A.}~\bibnamefont{Browaeys}},
  \bibinfo{author}{\bibfnamefont{J.}~\bibnamefont{Poupard}},
  \bibinfo{author}{\bibfnamefont{S.}~\bibnamefont{Nowak}},
  \bibinfo{author}{\bibfnamefont{D.}~\bibnamefont{Boiron}},
  \bibinfo{author}{\bibfnamefont{C.~I.} \bibnamefont{Westbrook}},
  \bibnamefont{and} \bibinfo{author}{\bibfnamefont{A.}~\bibnamefont{Aspect}},
  \bibinfo{journal}{Science} \textbf{\bibinfo{volume}{292}},
  \bibinfo{pages}{461} (\bibinfo{year}{2001}).

\bibitem[{\citenamefont{Zinner et~al.}(2000)\citenamefont{Zinner, Binnewies,
  Riehle, and Tiemann}}]{ZinBinRie00}
\bibinfo{author}{\bibfnamefont{G.}~\bibnamefont{Zinner}},
  \bibinfo{author}{\bibfnamefont{T.}~\bibnamefont{Binnewies}},
  \bibinfo{author}{\bibfnamefont{F.}~\bibnamefont{Riehle}}, \bibnamefont{and}
  \bibinfo{author}{\bibfnamefont{E.}~\bibnamefont{Tiemann}},
  \bibinfo{journal}{Phys.\ Rev.\ Lett.} \textbf{\bibinfo{volume}{85}},
  \bibinfo{pages}{2292} (\bibinfo{year}{2000}).

\bibitem[{\citenamefont{Katori et~al.}(1999)\citenamefont{Katori, Ido, Isoya,
  and {Kuwata-Gonokami}}}]{KatIdoIso99}
\bibinfo{author}{\bibfnamefont{H.}~\bibnamefont{Katori}},
  \bibinfo{author}{\bibfnamefont{T.}~\bibnamefont{Ido}},
  \bibinfo{author}{\bibfnamefont{Y.}~\bibnamefont{Isoya}}, \bibnamefont{and}
  \bibinfo{author}{\bibfnamefont{M.}~\bibnamefont{{Kuwata-Gonokami}}},
  \bibinfo{journal}{Phys.\ Rev.\ Lett.} \textbf{\bibinfo{volume}{82}},
  \bibinfo{pages}{1116} (\bibinfo{year}{1999}).

\bibitem[{\citenamefont{Dinneen et~al.}(1999)\citenamefont{Dinneen, Vogel,
  Hall, and Gallagher}}]{DinVogHal99}
\bibinfo{author}{\bibfnamefont{T.~P.} \bibnamefont{Dinneen}},
  \bibinfo{author}{\bibfnamefont{K.~R.} \bibnamefont{Vogel}},
  \bibinfo{author}{\bibfnamefont{J.~L.} \bibnamefont{Hall}}, \bibnamefont{and}
  \bibinfo{author}{\bibfnamefont{A.}~\bibnamefont{Gallagher}},
  \bibinfo{journal}{Phys. Rev. A} \textbf{\bibinfo{volume}{59}},
  \bibinfo{pages}{1216} (\bibinfo{year}{1999}).

\bibitem[{\citenamefont{Machholm et~al.}(1999)\citenamefont{Machholm, Julienne,
  and Suominen}}]{MacJulSuo99}
\bibinfo{author}{\bibfnamefont{M.}~\bibnamefont{Machholm}},
  \bibinfo{author}{\bibfnamefont{P.~S.} \bibnamefont{Julienne}},
  \bibnamefont{and} \bibinfo{author}{\bibfnamefont{K.~A.}
  \bibnamefont{Suominen}}, \bibinfo{journal}{Phys. Rev. A}
  \textbf{\bibinfo{volume}{59}}, \bibinfo{pages}{R4113} (\bibinfo{year}{1999}).

\bibitem[{\citenamefont{Wallis and Ertmer}(1989)}]{WalErt89}
\bibinfo{author}{\bibfnamefont{H.}~\bibnamefont{Wallis}} \bibnamefont{and}
  \bibinfo{author}{\bibfnamefont{W.}~\bibnamefont{Ertmer}},
  \bibinfo{journal}{J. Opt. Soc. Am. B} \textbf{\bibinfo{volume}{6}},
  \bibinfo{pages}{2211} (\bibinfo{year}{1989}).

\bibitem[{\citenamefont{Julienne}(2000)}]{Jul00}
\bibinfo{author}{\bibfnamefont{P.~S.} \bibnamefont{Julienne}}
  (\bibinfo{year}{2000}),
  \urlprefix\url{http://itamp.harvard.edu/alkaline-earth_atoms/julienne/online%
.html}.

\bibitem[{\citenamefont{Weiner et~al.}(1999)\citenamefont{Weiner, Bagnato,
  Zilio, and Julienne}}]{WeiBagZil99}
\bibinfo{author}{\bibfnamefont{J.}~\bibnamefont{Weiner}},
  \bibinfo{author}{\bibfnamefont{V.~S.} \bibnamefont{Bagnato}},
  \bibinfo{author}{\bibfnamefont{S.}~\bibnamefont{Zilio}}, \bibnamefont{and}
  \bibinfo{author}{\bibfnamefont{P.~S.} \bibnamefont{Julienne}},
  \bibinfo{journal}{Rev.\ Mod.\ Phys.} \textbf{\bibinfo{volume}{71}},
  \bibinfo{pages}{1} (\bibinfo{year}{1999}).

\bibitem[{\citenamefont{Dzuba et~al.}(1996{\natexlab{a}})\citenamefont{Dzuba,
  Flambaum, and Kozlov}}]{DzuFlaKoz96P}
\bibinfo{author}{\bibfnamefont{V.~A.} \bibnamefont{Dzuba}},
  \bibinfo{author}{\bibfnamefont{V.~V.} \bibnamefont{Flambaum}},
  \bibnamefont{and} \bibinfo{author}{\bibfnamefont{M.~G.}
  \bibnamefont{Kozlov}}, \bibinfo{journal}{Phys. Rev. A}
  \textbf{\bibinfo{volume}{54}}, \bibinfo{pages}{3948}
  (\bibinfo{year}{1996}{\natexlab{a}}).

\bibitem[{\citenamefont{Dzuba et~al.}(1998)\citenamefont{Dzuba, Kozlov, Porsev,
  and Flambaum}}]{DzuKozPor98}
\bibinfo{author}{\bibfnamefont{V.~A.} \bibnamefont{Dzuba}},
  \bibinfo{author}{\bibfnamefont{M.~G.} \bibnamefont{Kozlov}},
  \bibinfo{author}{\bibfnamefont{S.~G.} \bibnamefont{Porsev}},
  \bibnamefont{and} \bibinfo{author}{\bibfnamefont{V.~V.}
  \bibnamefont{Flambaum}}, \bibinfo{journal}{Zh. \ Eksp. \ Teor. \ Fiz.}
  \textbf{\bibinfo{volume}{114}}, \bibinfo{pages}{1636} (\bibinfo{year}{1998}),
  \bibinfo{note}{[Sov. \ Phys.--JETP {\bf 84} 461, (1997)]}.

\bibitem[{\citenamefont{Dalgarno and Davison}(1966)}]{DalDav66}
\bibinfo{author}{\bibfnamefont{A.}~\bibnamefont{Dalgarno}} \bibnamefont{and}
  \bibinfo{author}{\bibfnamefont{W.~D.} \bibnamefont{Davison}}, in
  \emph{\bibinfo{booktitle}{Adv.\ At.\ Mol.\ Phys.}}, edited by
  \bibinfo{editor}{\bibfnamefont{D.}~\bibnamefont{Bates}} \bibnamefont{and}
  \bibinfo{editor}{\bibfnamefont{I.}~\bibnamefont{Estermann}}
  (\bibinfo{publisher}{Academic Press}, \bibinfo{address}{New York},
  \bibinfo{year}{1966}), vol.~\bibinfo{volume}{2}, pp. \bibinfo{pages}{1--32}.

\bibitem[{\citenamefont{Derevianko et~al.}(1999)\citenamefont{Derevianko,
  Johnson, Safronova, and Babb}}]{DerJohSaf99}
\bibinfo{author}{\bibfnamefont{A.}~\bibnamefont{Derevianko}},
  \bibinfo{author}{\bibfnamefont{W.~R.} \bibnamefont{Johnson}},
  \bibinfo{author}{\bibfnamefont{M.~S.} \bibnamefont{Safronova}},
  \bibnamefont{and} \bibinfo{author}{\bibfnamefont{J.~F.} \bibnamefont{Babb}},
  \bibinfo{journal}{Phys.\ Rev.\ Lett.} \textbf{\bibinfo{volume}{82}},
  \bibinfo{pages}{3589} (\bibinfo{year}{1999}).

\bibitem[{\citenamefont{Derevianko et~al.}(2001)\citenamefont{Derevianko, Babb,
  and Dalgarno}}]{DerBabDal01}
\bibinfo{author}{\bibfnamefont{A.}~\bibnamefont{Derevianko}},
  \bibinfo{author}{\bibfnamefont{J.~F.} \bibnamefont{Babb}}, \bibnamefont{and}
  \bibinfo{author}{\bibfnamefont{A.}~\bibnamefont{Dalgarno}},
  \bibinfo{journal}{Phys.\ Rev.\ A} \textbf{\bibinfo{volume}{63}},
  \bibinfo{pages}{052704} (\bibinfo{year}{2001}).

\bibitem[{\citenamefont{Dzuba et~al.}(1996{\natexlab{b}})\citenamefont{Dzuba,
  Flambaum, and Kozlov}}]{DzuFlaKoz96Z}
\bibinfo{author}{\bibfnamefont{V.~A.} \bibnamefont{Dzuba}},
  \bibinfo{author}{\bibfnamefont{V.~V.} \bibnamefont{Flambaum}},
  \bibnamefont{and} \bibinfo{author}{\bibfnamefont{M.~G.}
  \bibnamefont{Kozlov}}, \bibinfo{journal}{Pis'ma Zh. Eks. Teor. Fiz.}
  \textbf{\bibinfo{volume}{63}}, \bibinfo{pages}{844}
  (\bibinfo{year}{1996}{\natexlab{b}}), \bibinfo{note}{[JETP \ Lett. {\bf 63},
  882 (1996)]}.

\bibitem[{\citenamefont{Porsev et~al.}(1999{\natexlab{a}})\citenamefont{Porsev,
  Rakhlina, and Kozlov}}]{PorRakKoz99J}
\bibinfo{author}{\bibfnamefont{S.~G.} \bibnamefont{Porsev}},
  \bibinfo{author}{\bibfnamefont{Y.~G.} \bibnamefont{Rakhlina}},
  \bibnamefont{and} \bibinfo{author}{\bibfnamefont{M.~G.}
  \bibnamefont{Kozlov}}, \bibinfo{journal}{J. Phys. B}
  \textbf{\bibinfo{volume}{32}}, \bibinfo{pages}{1113}
  (\bibinfo{year}{1999}{\natexlab{a}}).

\bibitem[{\citenamefont{Porsev et~al.}(1999{\natexlab{b}})\citenamefont{Porsev,
  Rakhlina, and Kozlov}}]{PorRakKoz99P}
\bibinfo{author}{\bibfnamefont{S.~G.} \bibnamefont{Porsev}},
  \bibinfo{author}{\bibfnamefont{Y.~G.} \bibnamefont{Rakhlina}},
  \bibnamefont{and} \bibinfo{author}{\bibfnamefont{M.~G.}
  \bibnamefont{Kozlov}}, \bibinfo{journal}{Phys. Rev. A}
  \textbf{\bibinfo{volume}{60}}, \bibinfo{pages}{2781}
  (\bibinfo{year}{1999}{\natexlab{b}}).

\bibitem[{\citenamefont{Kozlov and Porsev}(1999)}]{KozPor99E}
\bibinfo{author}{\bibfnamefont{M.~G.} \bibnamefont{Kozlov}} \bibnamefont{and}
  \bibinfo{author}{\bibfnamefont{S.~G.} \bibnamefont{Porsev}},
  \bibinfo{journal}{Eur. Phys. J. D} \textbf{\bibinfo{volume}{5}},
  \bibinfo{pages}{59} (\bibinfo{year}{1999}).

\bibitem[{\citenamefont{Porsev et~al.}(2000)\citenamefont{Porsev, Kozlov, and
  Rakhlina}}]{PorKozRak00Z}
\bibinfo{author}{\bibfnamefont{S.~G.} \bibnamefont{Porsev}},
  \bibinfo{author}{\bibfnamefont{M.~G.} \bibnamefont{Kozlov}},
  \bibnamefont{and} \bibinfo{author}{\bibfnamefont{Y.~G.}
  \bibnamefont{Rakhlina}}, \bibinfo{journal}{Pis'ma Zh. Eksp. Teor. Fiz.}
  \textbf{\bibinfo{volume}{72}}, \bibinfo{pages}{862} (\bibinfo{year}{2000}),
  \bibinfo{note}{[JETP \ Lett. {\bf 72} 595, (2000)]}.

\bibitem[{\citenamefont{Porsev et~al.}(2001)\citenamefont{Porsev, Kozlov,
  Rakhlina, and Derevianko}}]{PorKozRak01}
\bibinfo{author}{\bibfnamefont{S.~G.} \bibnamefont{Porsev}},
  \bibinfo{author}{\bibfnamefont{M.~G.} \bibnamefont{Kozlov}},
  \bibinfo{author}{\bibfnamefont{Y.~G.} \bibnamefont{Rakhlina}},
  \bibnamefont{and}
  \bibinfo{author}{\bibfnamefont{A.}~\bibnamefont{Derevianko}},
  \bibinfo{journal}{Phys. Rev. A} \textbf{\bibinfo{volume}{64}},
  \bibinfo{pages}{012508/1} (\bibinfo{year}{2001}).

\bibitem[{\citenamefont{Sternheimer}(1950)}]{Ste50}
\bibinfo{author}{\bibfnamefont{R.~M.} \bibnamefont{Sternheimer}},
  \bibinfo{journal}{Phys. Rev.} \textbf{\bibinfo{volume}{80}},
  \bibinfo{pages}{102} (\bibinfo{year}{1950}).

\bibitem[{\citenamefont{Dalgarno and Lewis}(1955)}]{DalLew55}
\bibinfo{author}{\bibfnamefont{A.}~\bibnamefont{Dalgarno}} \bibnamefont{and}
  \bibinfo{author}{\bibfnamefont{J.~T.} \bibnamefont{Lewis}},
  \bibinfo{journal}{Proc. Roy. Soc.} \textbf{\bibinfo{volume}{223}},
  \bibinfo{pages}{70} (\bibinfo{year}{1955}).

\bibitem[{\citenamefont{Johnson and Cheng}(1996)}]{JohChe96}
\bibinfo{author}{\bibfnamefont{W.~R.} \bibnamefont{Johnson}} \bibnamefont{and}
  \bibinfo{author}{\bibfnamefont{K.~T.} \bibnamefont{Cheng}},
  \bibinfo{journal}{Phys.\ Rev.\ A} \textbf{\bibinfo{volume}{53}},
  \bibinfo{pages}{1375} (\bibinfo{year}{1996}).

\bibitem[{\citenamefont{Moore}(1958)}]{Moo58}
\bibinfo{author}{\bibfnamefont{C.~E.} \bibnamefont{Moore}},
  \emph{\bibinfo{title}{Atomic energy levels}}, vol. \bibinfo{volume}{III}
  (\bibinfo{publisher}{National Bureau of Standards},
  \bibinfo{address}{Washington, D.C.}, \bibinfo{year}{1958}).

\bibitem[{\citenamefont{Bizzarri and Huber}(1990)}]{BizHub90}
\bibinfo{author}{\bibfnamefont{A.}~\bibnamefont{Bizzarri}} \bibnamefont{and}
  \bibinfo{author}{\bibfnamefont{M.~C.~E.} \bibnamefont{Huber}},
  \bibinfo{journal}{Phys. Rev. A} \textbf{\bibinfo{volume}{42}},
  \bibinfo{pages}{5422} (\bibinfo{year}{1990}).

\bibitem[{NIS()}]{NIST_ASD}
\emph{\bibinfo{title}{{NIST} atomic spectra database}},
  \urlprefix\url{http://physics.nist.gov/cgi-bin/AtData/main_asd}.

\bibitem[{\citenamefont{Johnson}(1988)}]{Joh88}
\bibinfo{author}{\bibfnamefont{W.~R.} \bibnamefont{Johnson}},
  \bibinfo{journal}{Adv. At. Mol. Phys.} \textbf{\bibinfo{volume}{25}},
  \bibinfo{pages}{375} (\bibinfo{year}{1988}).

\bibitem[{\citenamefont{Miller and Bederson}(1976)}]{MilBed76}
\bibinfo{author}{\bibfnamefont{T.~M.} \bibnamefont{Miller}} \bibnamefont{and}
  \bibinfo{author}{\bibfnamefont{B.}~\bibnamefont{Bederson}},
  \bibinfo{journal}{Phys. Rev. A} \textbf{\bibinfo{volume}{14}},
  \bibinfo{pages}{1572} (\bibinfo{year}{1976}).

\bibitem[{\citenamefont{Schwartz et~al.}(1974)\citenamefont{Schwartz, Miller,
  and Bederson}}]{SchMilBed74}
\bibinfo{author}{\bibfnamefont{H.~L.} \bibnamefont{Schwartz}},
  \bibinfo{author}{\bibfnamefont{T.~M.} \bibnamefont{Miller}},
  \bibnamefont{and} \bibinfo{author}{\bibfnamefont{B.}~\bibnamefont{Bederson}},
  \bibinfo{journal}{Phys. Rev. A} \textbf{\bibinfo{volume}{10}},
  \bibinfo{pages}{1924} (\bibinfo{year}{1974}).

\bibitem[{\citenamefont{Stanton}(1994)}]{Sta94}
\bibinfo{author}{\bibfnamefont{J.~F.} \bibnamefont{Stanton}},
  \bibinfo{journal}{Phys. Rev. A} \textbf{\bibinfo{volume}{49}},
  \bibinfo{pages}{1698} (\bibinfo{year}{1994}).

\bibitem[{\citenamefont{Standard and Certain}(1985)}]{StaCer85}
\bibinfo{author}{\bibfnamefont{J.~M.} \bibnamefont{Standard}} \bibnamefont{and}
  \bibinfo{author}{\bibfnamefont{P.~R.} \bibnamefont{Certain}},
  \bibinfo{journal}{J. Chem. Phys.} \textbf{\bibinfo{volume}{83}},
  \bibinfo{pages}{3002} (\bibinfo{year}{1985}).

\bibitem[{\citenamefont{Maeder and Kutzelnigg}(1979)}]{MaeKut79}
\bibinfo{author}{\bibfnamefont{F.}~\bibnamefont{Maeder}} \bibnamefont{and}
  \bibinfo{author}{\bibfnamefont{W.}~\bibnamefont{Kutzelnigg}},
  \bibinfo{journal}{Chem.\ Phys.} \textbf{\bibinfo{volume}{42}},
  \bibinfo{pages}{95} (\bibinfo{year}{1979}).

\bibitem[{\citenamefont{Amusia and Cherepkov}(1975)}]{AmuChe75}
\bibinfo{author}{\bibfnamefont{M.~Y.} \bibnamefont{Amusia}} \bibnamefont{and}
  \bibinfo{author}{\bibfnamefont{N.~A.} \bibnamefont{Cherepkov}},
  \bibinfo{journal}{Case Studies in Atomic Physics}
  \textbf{\bibinfo{volume}{5}}, \bibinfo{pages}{47} (\bibinfo{year}{1975}).

\bibitem[{\citenamefont{Stwalley}(1971)}]{Stw71}
\bibinfo{author}{\bibfnamefont{W.~C.} \bibnamefont{Stwalley}},
  \bibinfo{journal}{J. Chem. Phys.} \textbf{\bibinfo{volume}{54}},
  \bibinfo{pages}{4517} (\bibinfo{year}{1971}).

\bibitem[{\citenamefont{Derevianko}()}]{Der01a}
\bibinfo{author}{\bibfnamefont{A.}~\bibnamefont{Derevianko}},
  \bibinfo{note}{e-print:physics/0108041}.

\bibitem[{\citenamefont{Leo et~al.}(2000)\citenamefont{Leo, Williams, and
  Julienne}}]{LeoWilJul00}
\bibinfo{author}{\bibfnamefont{P.~J.} \bibnamefont{Leo}},
  \bibinfo{author}{\bibfnamefont{C.~J.} \bibnamefont{Williams}},
  \bibnamefont{and} \bibinfo{author}{\bibfnamefont{P.~S.}
  \bibnamefont{Julienne}}, \bibinfo{journal}{Phys.\ Rev.\ Lett.}
  \textbf{\bibinfo{volume}{85}}, \bibinfo{pages}{2721} (\bibinfo{year}{2000}).

\end{thebibliography}

\end{document}